\documentclass[conference]{IEEEtran}

\usepackage[cmex10]{amsmath}
\usepackage{amssymb}
\usepackage{bm}
\usepackage{cite}
\usepackage{tikz}
\usetikzlibrary{shapes,arrows}
\usepackage{subfigure}
\usepackage{algorithm}
\usepackage[noend]{algorithmic}
\usepackage{bm}

\newtheorem{theorem}{Theorem}
\newtheorem{lemma}{Lemma}

\begin{document}

\title{Adaptive Linear Programming Decoding of \\ Polar Codes}

\author{\IEEEauthorblockN{Veeresh Taranalli and Paul H.\@ Siegel}
\IEEEauthorblockA{University of California, San Diego, La Jolla, CA 92093, USA\\
Email: \{vtaranalli, psiegel\}@ucsd.edu}}

\maketitle

%% ========================================================
%%  Abstract
%% ========================================================
\begin{abstract}
Polar codes are high density parity check codes and hence the sparse factor graph, instead of the parity check matrix, has been used to practically represent an LP polytope for LP decoding. Although LP decoding on this polytope has the ML-certificate property, it performs poorly over a BAWGN channel. In this paper, we propose modifications to adaptive cut generation based LP decoding techniques and apply the modified-adaptive LP decoder to short blocklength polar codes over a BAWGN channel. The proposed decoder provides significant FER performance gain compared to the previously proposed LP decoder and its performance approaches that of ML decoding at high SNRs. We also present an algorithm to obtain a smaller factor graph from the original sparse factor graph of a polar code. This reduced factor graph  preserves the small check node degrees needed to represent the LP polytope in practice. We show that the fundamental polytope of the reduced factor graph can be obtained from the projection of the polytope represented by the original sparse factor graph and the frozen bit information. Thus, the LP decoding time complexity is decreased without changing the FER performance by using the reduced factor graph representation.
\end{abstract}

%% ========================================================
%% Introduction
%% ========================================================
\section{Introduction}
\label{sec:intro}
Polar codes, first introduced in \cite{arikan2009}, were shown to be capacity-achieving for binary input memoryless output symmetric channels. However, their performance on a binary additive white gaussian noise channel (BAWGNC) with successive cancellation (SC) decoding is unimpressive at practical blocklengths. Thus, improving their performance either by improved decoding algorithms or modified constructions of polar codes has been a recent topic in coding theory. The most notable improvement in error rate performance was observed using the successive cancellation list (SC-List) decoding algorithm proposed in~\cite{tal_list_2011}. Alternatively, a cyclic redundancy check (CRC) concatenated polar code with SC-List decoding was also shown to improve the performance significantly~\cite{tal_list_2011}.

Linear Programming (LP) decoding has been a research topic in coding theory and is attractive mainly because of its maximum likelihood (ML)-certificate property~\cite{feldman2005}. It was introduced for polar codes in~\cite{goela2010} where the sparse factor graph was used to represent the LP polytope instead of the high density parity check matrix. For polar codes over a binary erasure channel (BEC), it was shown that LP decoding achieves capacity and also outperforms SC decoding at finite blocklengths~\cite{goela2010}. However, for a BAWGNC, the LP decoder in~\cite{goela2010} is suboptimal and performs very poorly.  

Adaptive LP decoding techniques were proposed in~\cite{taghaviJ2008, zhang2012} to improve the decoding time complexity as well as the error rate performance. Based on these techniques, we propose modifications to the adaptive cut generation based LP decoder in~\cite{zhang2012} that significantly improve its error rate performance for polar codes over a BAWGNC. We then present an algorithm to obtain a smaller factor graph representation of a polar code called the \emph{reduced factor graph}, which decreases the representation complexity of the fundamental polytope and hence improves the decoding time complexity of the modified-adaptive LP decoder.       

In Section~\ref{sec:lp_decoding_polar}, we review the LP decoding of polar codes proposed in~\cite{goela2010}. In Section~\ref{sec:alp_decoding}, we review the adaptive LP decoding techniques~\cite{taghaviJ2008, zhang2012} and describe the proposed modified-adaptive LP decoder for short blocklength polar codes, along with simulation results. In Section~\ref{sec:reduced_fg}, we present the algorithm for reducing a polar code sparse factor graph. 

%% =======================================================
%% LP Decoding of Polar Codes
%% =======================================================
\section{LP Decoding of Polar Codes}
\label{sec:lp_decoding_polar}
Consider a binary linear code $\mathcal{C}_l$ of length $N$ and rate $r = \frac{k}{N}$, where $k < N$ is the number of information bits in a codeword. Let $\mathbf{H}$ denote a parity check matrix for $\mathcal{C}_l$. Suppose a codeword $\mathbf{x} \in \mathcal{C}_l$ is transmitted over a binary input memoryless output symmetric channel and $\mathbf{y}$ is the received vector. ML decoding is equivalent to solving the optimization problem \cite{feldman2005}:
\begin{IEEEeqnarray}{C}
	\label{basic_ilp}
	\textrm{minimize}~ \bm{\gamma}^{T}\mathbf{x} ~~
	\textrm{subject to}~\mathbf{x} \in \mathcal{C}_l
\end{IEEEeqnarray}
where $ x_{i} \in \{0, 1\}, i \in {1,\ldots,N} $ and $\bm{\gamma}$ is the vector of log-likelihood ratios (LLR) defined as 
\begin{IEEEeqnarray}{C}
\gamma_{i} = \textrm{log} \Bigg( \frac{\textrm{Pr}(y_i|x_i = 0)}{\textrm{Pr}(y_i|x_i = 1)} \Bigg).
\end{IEEEeqnarray}
In \cite{feldman2005}, the ML decoding problem (\ref{basic_ilp}) was relaxed to a linear programming (LP) problem, where the relaxed polytope, also known as the fundamental polytope $\mathcal{Q}$ has both integral and non-integral vertices. The polytope $\mathcal{Q}$ is defined by linear inequalities, also referred to as constraints, generated from each row $j$ of the parity-check matrix $\mathbf{H}$, given by
\begin{IEEEeqnarray}{C}
\label{feldman_polytope_inequalities}
\sum\limits_{i \in V} x_{i} - \sum\limits_{i\in\mathcal{N}(j)\backslash V} x_{i} \leq |V| - 1\\ 
\nonumber \forall~V \subseteq \mathcal{N}(j)~~\textrm{s.t.}~~|V|~~\textrm{is odd}
\end{IEEEeqnarray}
where $\mathcal{N}(j)$ is the support of the row $j$ of $\mathbf{H}$. This polytope $\mathcal{Q}$ has the ML-certificate property which guarantees that an integral solution of the LP problem would be a valid ML codeword. The number of constraints needed to define the polytope $\mathcal{Q}$ is exponential in the maximum parity-check degree of $\mathbf{H}$, i.e., $O(2^{\max\limits_{j}|\mathcal{N}(j)|}).$ Let a polar code $\mathcal{C}$ be constructed using the channel polarization transform of length $N = 2^{m}$ proposed in \cite{arikan2009}, which is denoted by a matrix $\mathbf{G_N}$, where $\mathbf{G_N} = \mathbf{B_N}\mathbf{G_{2}}^{\otimes m}$, $\mathbf{G_2} = \bigl [ \begin{smallmatrix} 1 & 0 \\ 1 & 1 \end{smallmatrix} \bigr ] $, the operator $\otimes m$ represents the $m$-times Kronecker product of $\mathbf{G_{2}}$, and $\mathbf{B_N}$ is the bit-reversal permutation matrix defined in \cite{arikan2009}. Assuming all the $N-k$ frozen (non-information) bits in $\mathcal{C}$ are set to $0$, a parity check matrix $\mathbf{H}$ for $\mathcal{C}$ can be constructed by selecting the columns corresponding to the frozen bit indices in $\mathbf{G_{N}}$ as the parity checks~\cite{goela2010}. Thus, $\mathbf{H}$ consists of high density rows with the maximum possible parity-check degree being $N$. Hence, the number of constraints needed to define the polytope $\mathcal{Q}$ as per (\ref{feldman_polytope_inequalities}) is $O(2^{N-1})$. This is clearly impractical for all but very short length polar codes. It was also shown in \cite{goela2010} that LP decoding on the fundamental polytope $\mathcal{Q}$ will fail for a BEC($\epsilon$), binary symmetric channel BSC($p$) or a BAWGNC($\sigma$), even if the polytope $\mathcal{Q}$ could be represented in practice.

A sparse factor graph representation with $O(N\textrm{log}N)$ auxiliary variable nodes was proposed in~\cite{arikan2009} for the polar code $\mathcal{C}$. An example sparse factor graph for $N~=~8$ is shown in Fig.~\ref{fig:sparse_fgraph_8}. It is easy to see that there are only degree-3 or degree-2 check nodes in the sparse factor graph. Let $\mathbf{H}_\mathcal{P}$ denote the adjacency matrix of the sparse factor graph where the rows and columns represent the check and variable nodes, respectively. The LP polytope $\mathcal{P}$ is defined as the intersection of local minimal convex polytopes of each row (parity check) in $\mathbf{H}_\mathcal{P}$ and the set of cutting planes corresponding to the frozen column indices (variable nodes) in $\mathbf{H}_\mathcal{P}$~\cite{goela2010}.

Using the polytope $\mathcal{P}$, an LP decoder for the polar code $\mathcal{C}$ as proposed in \cite{goela2010} is given by
\begin{IEEEeqnarray}{C}
\label{polar_lp}
	\nonumber
	\textrm{minimize}~ \bm{\gamma}^{T}\mathbf{\bar{x}} \\
	\textrm{subject to}~\mathbf{x} \in \mathcal{P} \subseteq [0, 1]^{N(1 + \textrm{log}N)}
\end{IEEEeqnarray} 
where $\mathbf{\bar{x}}$ is defined as the first $N$ component subset of $\mathbf{x}$ and corresponds to the codeword variable nodes, $\bar{x}_{i}~=~ x_i~\forall~i~\in~\{1,\ldots,N\}$. Similarly, the projection of polytope $\mathcal{P}$ is defined \cite{goela2010} as
\begin{IEEEeqnarray}{C}
	\mathcal{\bar{P}} = \{\mathbf{\bar{x}}~\in~[0, 1]^{N}~|~\exists~\mathbf{\hat{x}}~s.t.~(\mathbf{\bar{x}}, \mathbf{\hat{x}})~\in \mathcal{P}\}
\end{IEEEeqnarray} 

It was shown that if the projection $\mathbf{\bar{x}}$ (on $\mathcal{\bar{P}}$) of the LP decoder output vector $\mathbf{x}$ in (\ref{polar_lp}) is integral then it is guaranteed to be the ML codeword i.e., LP decoding on the polytope $\mathcal{P}$ as defined in (\ref{polar_lp}) has the ML-certificate property (Lemma 3 in \cite{goela2010}). 
It was also shown (Theorem~1~in~\cite{goela2010}) that the projection $\mathcal{\bar{P}}$ of polytope $\mathcal{P}$ is tighter than the fundamental polytope $\mathcal{Q}$. 

% Define block styles
\tikzstyle{var} = [circle, fill, minimum size=3pt,inner sep=0pt, outer sep=0pt]
\tikzstyle{check} = [rectangle, fill, minimum size=4pt,inner sep=0pt, outer sep=0pt]
\tikzstyle{every node}=[font=\footnotesize]
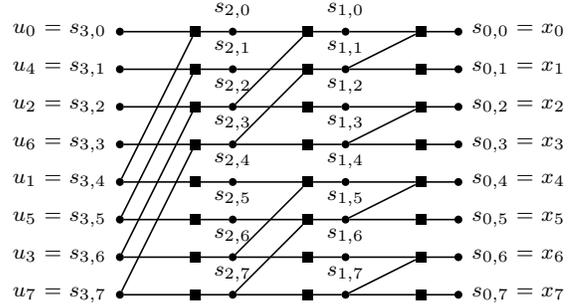
\begin{figure}[!ht]
\centering
\begin{tikzpicture}[yscale=0.5, xscale=1.0, node distance=0.3cm, auto, semithick]
    \node[var] (Nvar-3-0) [label=left:$u_{0}~\mathrm{=}~s_{3, 0}$] at (0,0) {};
    \node[var] (Nvar-3-1) [label=left:$u_{4}~\mathrm{=}~s_{3, 1}$] at (0,-1) {};
    \node[var] (Nvar-3-2) [label=left:$u_{2}~\mathrm{=}~s_{3, 2}$] at (0,-2) {};
    \node[var] (Nvar-3-3) [label=left:$u_{6}~\mathrm{=}~s_{3, 3}$] at (0,-3) {};
    \node[var] (Nvar-3-4) [label=left:$u_{1}~\mathrm{=}~s_{3, 4}$] at (0,-4) {};
    \node[var] (Nvar-3-5) [label=left:$u_{5}~\mathrm{=}~s_{3, 5}$] at (0,-5) {};
    \node[var] (Nvar-3-6) [label=left:$u_{3}~\mathrm{=}~s_{3, 6}$] at (0,-6) {};
    \node[var] (Nvar-3-7) [label=left:$u_{7}~\mathrm{=}~s_{3, 7}$] at (0,-7) {};   
    \foreach \y in {0,...,7}
        \node[check] (Ncheck-2-\y) at (1,-\y) {};
	\foreach \y in {0,...,7}
		\path (Nvar-3-\y) edge[-] (Ncheck-2-\y);
	\foreach \y / \x in {4/0, 5/1, 6/2, 7/3}
		\path (Nvar-3-\y) edge[-] (Ncheck-2-\x);
    
	\foreach \y in {0,...,7}
		\pgfmathtruncatemacro\result{\y + 17}
        \node[var] (Nvar-2-\y) [label=above:$s_{2, \y}$] at (1.5,-\y) {};
    \foreach \y in {0,...,7}
    	\path (Ncheck-2-\y) edge[-] (Nvar-2-\y);
    	
   	\foreach \y in {0,...,7}
        \node[check] (Ncheck-1-\y) at (2.5,-\y) {};
    \foreach \y in {0,...,7}
		\path (Nvar-2-\y) edge[-] (Ncheck-1-\y);   
	\foreach \y / \x in {2/0, 3/1, 6/4, 7/5}
	 	\path (Nvar-2-\y) edge[-] (Ncheck-1-\x);
	 
	 \foreach \y in {0,...,7}
	 	\pgfmathtruncatemacro\result{\y + 9}
        \node[var] (Nvar-1-\y) [label=above:$s_{1, \y}$] at (3.0,-\y) {};
    \foreach \y in {0,...,7}
    	\path (Ncheck-1-\y) edge[-] (Nvar-1-\y);
    	
   	\foreach \y in {0,...,7}
        \node[check] (Ncheck-0-\y) at (4.0,-\y) {};
    \foreach \y in {0,...,7}
		\path (Nvar-1-\y) edge[-] (Ncheck-0-\y);   
	\foreach \y / \x in {1/0, 3/2, 5/4, 7/6}
	 	\path (Nvar-1-\y) edge[-] (Ncheck-0-\x);
	
	\foreach \y in {0,...,7}
		\pgfmathtruncatemacro\result{\y + 1}
		\node[var] (Nvar-0-\y)  [label=right:$s_{0, \y}~\mathrm{=}~x_{\y}$] at (4.5, -\y) {};
	\foreach \y in {0,...,7}
		\path (Nvar-0-\y) edge[-] (Ncheck-0-\y);   	 
              
\end{tikzpicture}
\caption{Sparse factor graph representation of a length-8 polar code.}
\label{fig:sparse_fgraph_8}
\end{figure}

\setlength{\textfloatsep}{0.87em}% Remove \textfloatsep

%% =====================================================
%% Adaptive LP Decoding of Polar Codes
%% =====================================================
\section{Adaptive LP Decoding of Polar Codes}
\label{sec:alp_decoding}

\subsection{Adaptive LP Decoding of a Binary Linear Code}
\label{subsec:alp_decoding_linear_code}
An Adaptive LP (ALP) decoder for binary linear codes solves a sequence of LP decoding problems with the addition of intelligently chosen constraints called \emph{cuts} at every iteration~\cite{taghaviJ2008}. A cut at a point $\mathbf{x}~\in~[0, 1]^{N}$ is a violated constraint at $\mathbf{x}$ derived from a check node. An ALP decoder starts by solving the initial LP with the constraints 
\begin{IEEEeqnarray}{C} 
x_i \geq 0 ~~ \textrm{if} ~~ \gamma_i \geq 0 ;~~~~~~
x_i \leq 1 ~~ \textrm{if} ~~ \gamma_i < 0 
\label{eq:box_constraints}
\end{IEEEeqnarray}
The solution of this initial LP coincides with the output of a hard decision decoding of the received LLR values. The ALP decoder then searches constraints from all parity checks to find cuts, adds the cuts to the LP and solves the resulting LP\@. This procedure is repeated until an integer solution is obtained or no further cuts can be found. Violated constraints or cuts can also be generated using redundant parity checks (RPCs). RPCs are obtained by the modulo-2 addition of parity checks in the parity check matrix of the code. The addition of cuts from RPCs during the LP decoding iterations can only tighten the LP polytope and hence can only improve the error rate performance. 
In~\cite{zhang2012}, efficient algorithms to perform the cut-search on parity checks and to find cut-inducing RPCs were proposed. Based on these algorithms, an adaptive cut generation based LP (ACG-ALP) decoder (Algorithm~2 in~\cite{zhang2012}) was proposed. Next, we present modifications to the ACG-ALP decoder which make it suitable for decoding polar codes.  

\subsection{Modified ACG-ALP Decoder for Polar Codes}
\label{subsec:acg_alp_polar}
A polar code can be defined using the sparse factor graph ($\mathbf{H}_\mathcal{P}$) with the frozen bit information or the parity check matrix ($\mathbf{H}$). 
The availability of these two representations motivates the idea of modifying the ACG-ALP decoder (Algorithm~2 in~\cite{zhang2012}) to improve its performance when compared to a LP decoder. 
The ACG-ALP decoder uses the parity check matrix to generate constraints and cuts. RPCs and cuts from these RPCs are also derived from the parity check matrix by the ACG-ALP decoder. 
Based on these observations, we investigate four ways of using the sparse factor graph and the parity check matrix representations in the ACG-ALP decoder: 
\begin{enumerate}
\item[1.] Use the unmodified ACG-ALP decoder with the
parity check matrix $\mathbf{H}$.
\item[2.] Use $\mathbf{H}_\mathcal{P}$ as the parity check matrix in the ACG-ALP decoder.
Add the frozen bit constraints to the inital-LP.\@
\item[3.] Initialize the ACG-ALP decoder with the polytope $\mathcal{P}$ and generate subsequent cut-inducing RPCs from $\mathbf{H}_\mathcal{P}$.
\item[4.] Initialize the ACG-ALP decoder with the polytope $\mathcal{P}$ and generate subsequent cut-inducing RPCs from $\mathbf{H}$. 
\end{enumerate}

\begin{figure}[!ht]
\includegraphics[width=0.48\textwidth]{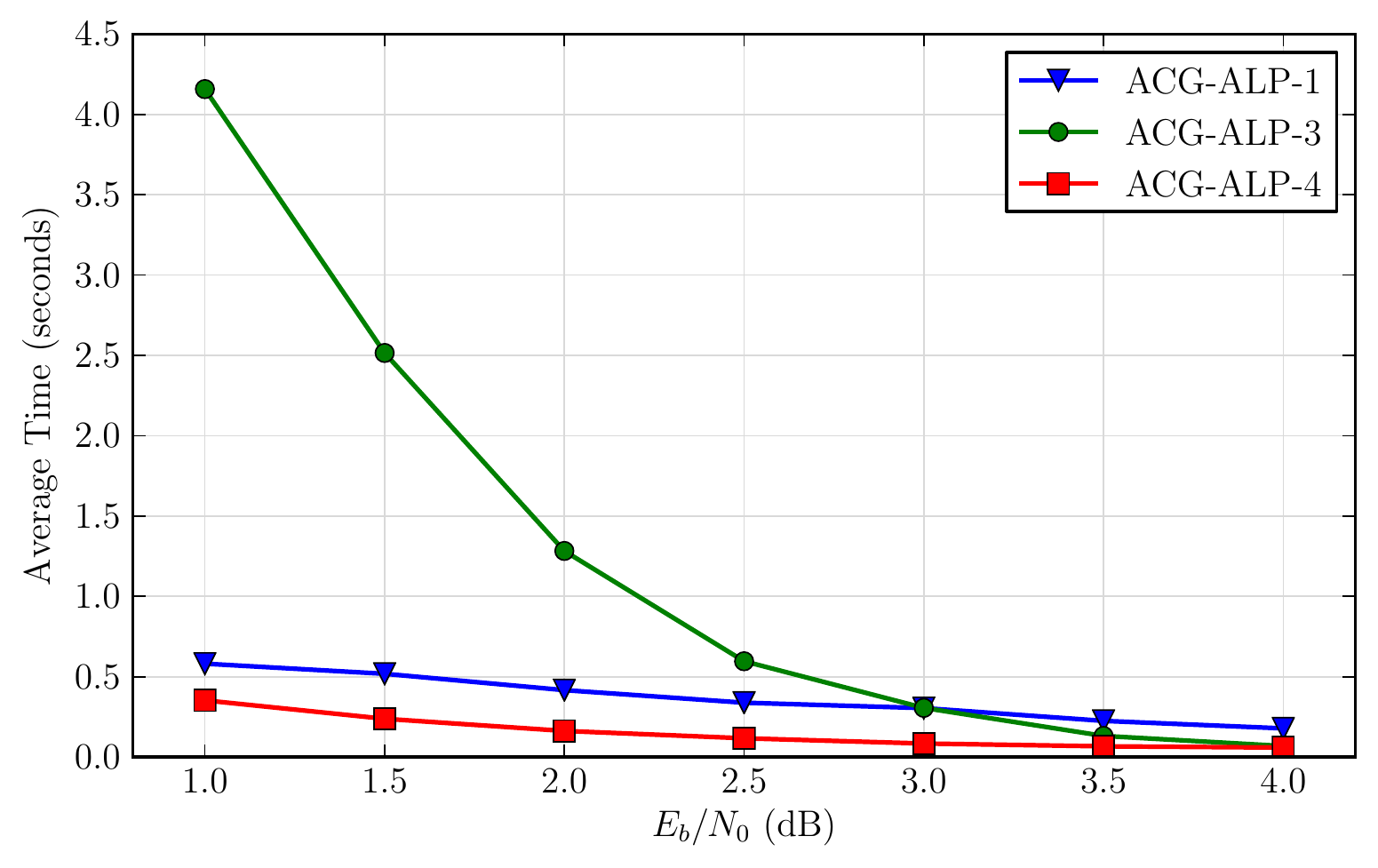}
\caption{Average time for decoding one codeword of a (64, 32) polar code over a BAWGN channel with ACG-ALP decoding.}
\label{fig:avgtime_acgalp_14}
\end{figure}

Note that the ACG-ALP decoders 1 and 2 do not use constraints from parity checks in defining the initial LP\@. Hence, these decoders are expected to have a larger average decoding time complexity compared to the ACG-ALP decoders 3 and 4, as shown in Fig.~\ref{fig:avgtime_acgalp_14}. Due to its large decoding time complexity, the simulation results corresponding to the ACG-ALP~decoder~2 could not be obtained. We have empirically observed that the other three ACG-ALP decoders (1,~3,~4) perform equally well in terms of the frame error rate (FER) performance but the ACG-ALP~decoder~4 has the smallest average decoding time complexity. Hence, we select this modified decoder for decoding polar codes (ACG-ALP-Polar) as shown in Algorithm~\ref{algo:ACG_ALP_polar_algo}.

\begin{algorithm}[h]
\caption{ACG-ALP decoding algorithm for Polar codes}
\label{algo:ACG_ALP_polar_algo}
\algsetup{
linenosize=\small
}
\begin{algorithmic}[1]
\REQUIRE $\bm{\gamma}$, $\mathbf{H}_\mathcal{P}$, frozen bit indices, $\mathbf{H}$
\ENSURE Optimal solution of the current LP problem
\STATE Initialize the LP problem with the constraints obtained from $\mathbf{H}_\mathcal{P}$ and frozen bit information.
\STATE Solve the current LP problem to get the solution~$\mathbf{x}^{*}$.
\IF{$\mathbf{x}^{*}$ is nonintegral}
	\STATE Construct cut-inducing RPC matrix $\tilde{\mathbf{H}}$ from $\mathbf{H}$~\cite{zhang2012}.
	\STATE Apply the cut-search algorithm (Algorithm~1 in~\cite{zhang2012}) to each row of $\tilde{\mathbf{H}}$.
\ENDIF
\IF{No cut is found}
	\STATE Terminate.
\ELSE
	\STATE Add the cuts found to the LP problem, go to line 2.
\ENDIF
\end{algorithmic}
\end{algorithm}

\subsection{Simulation Results}
\label{subsec:acg_alp_polar_sim_results}
\begin{figure}
\includegraphics[width=0.5\textwidth]{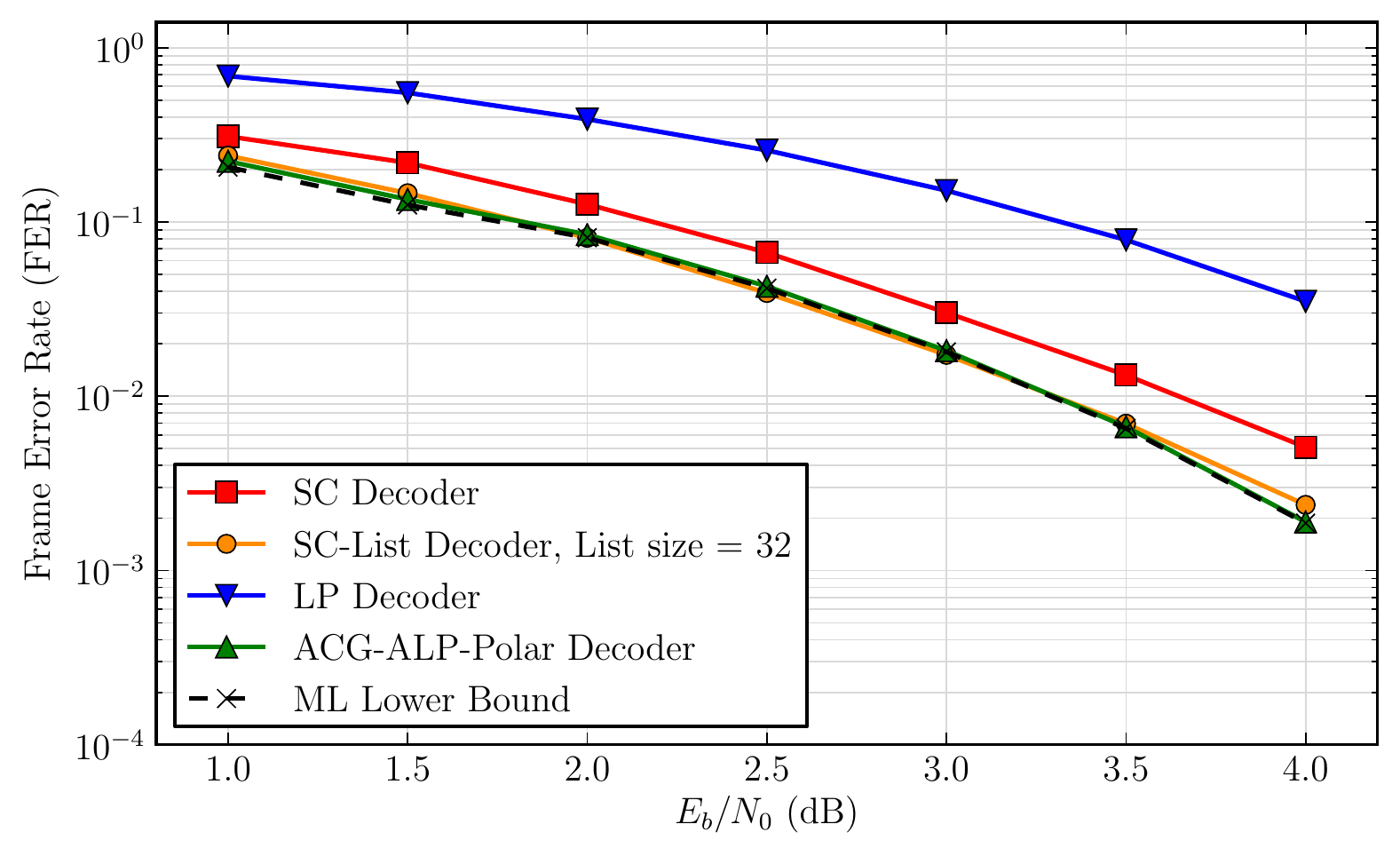}
\caption{FER performance of a (64, 32) polar code over a BAWGN channel.}
\label{fig:acg_alp_ofg_64}
\end{figure}

\begin{figure}
\includegraphics[width=0.5\textwidth]{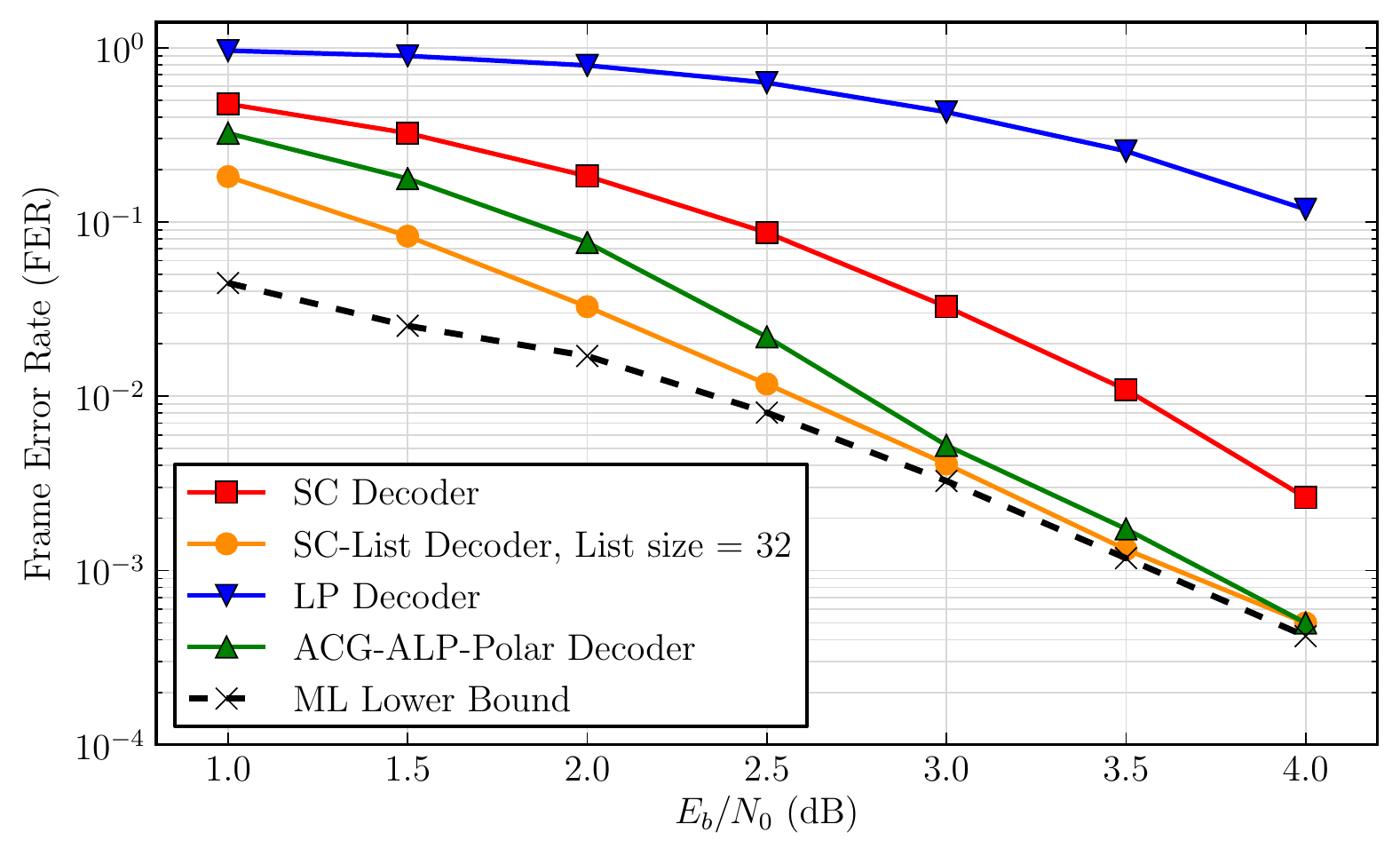}
\caption{FER performance of a (128, 64) polar code over a BAWGN channel.}
\label{fig:acg_alp_ofg_128}
\end{figure}

Fig.~\ref{fig:acg_alp_ofg_64} and Fig.~\ref{fig:acg_alp_ofg_128} show the FER performance over a BAWGNC of rate-0.5 length-64 and length-128 polar codes, respectively. The performance of the proposed ACG-ALP-Polar decoder is compared with the previously proposed LP decoder for polar codes~\cite{goela2010}, the SC and the SC-List decoders~\cite{tal_list_2011}. We choose a list size~=~32 for the SC-List decoder (SC-List-32) as it is known to have performance close to the ML lower bound~\cite{tal_list_2011}. The polar codes are constructed using the bit channel degrading merge algorithm presented in~\cite{tal_vardy_itt_2013} optimized for a BAWGN channel with signal-to-noise ratio, SNR~$(E_s/N_0)$~=~3.0~dB.\@ The proposed decoder uses the LP solver in the CVXOPT package~\cite{cvxopt}. A total of 200 frame errors were recorded at each $E_b/N_0$ point. We also show an ML lower bound obtained using the ML-certificate property of the proposed decoder. The ACG-ALP-Polar decoder shows a significant improvement in performance compared to the LP decoder for both polar codes. For the length-64 polar code, the ACG-ALP-Polar decoder performance is very close to the ML lower bound. However, for the length-128 polar code, there is a performance gap in the lower SNR region. The ACG-ALP-Polar decoder performs better than the SC decoder for both polar codes studied. Compared to the SC-List-32 decoder, the ACG-ALP-Polar decoder performs equally well for the $(64,~32)$ polar code while its performance is worse in the low SNR region for the $(128,~64)$ polar code. Although the ACG-ALP-Polar decoder shows promise in FER performance, we have empirically observed that its decoding time complexity is larger than that of the SC-List-32 decoder especially in the low SNR region and is prohibitively high for moderate to long blocklength polar codes.

\setlength{\textfloatsep}{2pt}

%% ========================================================
%% Polar Code Sparse Factor Graph Reduction
%% ========================================================
\section{Polar Code Sparse Factor Graph Reduction}
\label{sec:reduced_fg}
Due to its recursive structure the polar code sparse factor graph has some redundant variable nodes connected to \mbox{degree-2} check nodes. We use this redundancy and the frozen bit information to propose an algorithm for reducing the number of constraints needed to represent the sparse factor graph based LP polytope $\mathcal{P}$. Simulation results show that the new reduced factor graph representation can be used in the ACG-ALP-Polar decoder to achieve an improvement in the time complexity.

\subsection{Polar Code Sparse Factor Graph Reduction Algorithm}
The sparse factor graph, the set of frozen bit indices and their values define the polar code. The set of frozen bit indices is obtained using a polar code construction technique~\cite{tal_vardy_itt_2013}. We assume that all frozen bits are set to $0$.

\tikzstyle{var} = [circle, fill, minimum size=4pt,inner sep=0pt, outer sep=0pt]
\tikzstyle{check} = [rectangle, fill, minimum size=4pt,inner sep=0pt, outer sep=0pt]
\tikzstyle{selected check} = [check, fill=blue, minimum size=4.5pt]
\tikzstyle{selected var} = [var, fill=blue, minimum size=4.5pt]
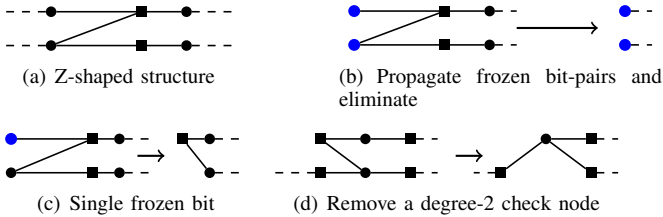
\begin{figure}[!ht]
\subfigure[Z-shaped structure]{
\label{fig:fg_reduction_a}
\begin{tikzpicture}[yscale=0.45, xscale=1.2, node distance=0.2cm, auto, semithick]
\node[var] (Nvar-1-0) at (0.0, 0) {};
\node[var] (Nvar-1-1) at (0.0, -1) {};
\node[var] (Nvar-0-0) at (1.5, 0) {};
\node[var] (Nvar-0-1) at (1.5, -1) {};
\node[check] (Ncheck-0-0) at (1.0, 0) {};
\node[check] (Ncheck-0-1) at (1.0, -1) {};
\path (Nvar-0-0) edge[-] (Ncheck-0-0);
\path (Nvar-0-1) edge[-] (Ncheck-0-1);
\path (Nvar-1-0) edge[-] (Ncheck-0-0);
\path (Nvar-1-1) edge[-] (Ncheck-0-1);
\path (Nvar-1-1) edge[-] (Ncheck-0-0);
\draw[-, dashed] (Nvar-0-0) -- (2.0, 0);
\draw[-, dashed] (Nvar-0-1) -- (2.0, -1);
\draw[-, dashed] (Nvar-1-0) -- (-0.5, 0);
\draw[-, dashed] (Nvar-1-1) -- (-0.5, -1);
\end{tikzpicture}
}
\subfigure[Propagate frozen bit-pairs and eliminate]{
\label{fig:fg_reduction_b}
\begin{tikzpicture}[yscale=0.45, xscale=1.2, node distance=0.2cm, auto, semithick]
\node[selected var] (Nvar-1-0) at (0.0, 0) {};
\node[selected var] (Nvar-1-1) at (0.0, -1) {};
\node[var] (Nvar-0-0) at (1.5, 0) {};
\node[var] (Nvar-0-1) at (1.5, -1) {};
\node[check] (Ncheck-0-0) at (1.0, 0) {};
\node[check] (Ncheck-0-1) at (1.0, -1) {};
\path (Nvar-0-0) edge[-] (Ncheck-0-0);
\path (Nvar-0-1) edge[-] (Ncheck-0-1);
\path (Nvar-1-0) edge[-] (Ncheck-0-0);
\path (Nvar-1-1) edge[-] (Ncheck-0-1);
\path (Nvar-1-1) edge[-] (Ncheck-0-0);
\draw[->, thick] (1.8, -0.5) -- (2.7, -0.5); 
\node[selected var] (Nvar-2-0) at (3.0, 0) {};
\node[selected var] (Nvar-2-1) at (3.0, -1) {};
\draw[-, dashed] (Nvar-0-0) -- (1.8, 0);
\draw[-, dashed] (Nvar-0-1) -- (1.8, -1);
\draw[-, dashed] (Nvar-2-0) -- (3.3, 0);
\draw[-, dashed] (Nvar-2-1) -- (3.3, -1);
\end{tikzpicture}
}
\subfigure[Single frozen bit]{
\label{fig:fg_reduction_c}
\begin{tikzpicture}[yscale=0.45, xscale=1.2, node distance=0.2cm, auto, semithick]
\node[selected var] (Nvar-1-0) at (0.0, 0) {};
\node[var] (Nvar-1-1) at (0.0, -1) {};
\node[var] (Nvar-0-0) at (1.2, 0) {};
\node[var] (Nvar-0-1) at (1.2, -1) {};
\node[check] (Ncheck-0-0) at (0.9, 0) {};
\node[check] (Ncheck-0-1) at (0.9, -1) {};
\path (Nvar-0-0) edge[-] (Ncheck-0-0);
\path (Nvar-0-1) edge[-] (Ncheck-0-1);
\path (Nvar-1-0) edge[-] (Ncheck-0-0);
\path (Nvar-1-1) edge[-] (Ncheck-0-1);
\path (Nvar-1-1) edge[-] (Ncheck-0-0);
\draw[->, thick] (1.4, -0.5) -- (1.7, -0.5); 
\node[var] (Nvar-2-0) at (2.2, 0) {};
\node[var] (Nvar-2-1) at (2.2, -1) {};
\node[check] (Ncheck-2-0) at (1.9, 0) {};
\path (Nvar-2-0) edge[-] (Ncheck-2-0);
\path (Nvar-2-1) edge[-] (Ncheck-2-0);
\draw[-, dashed] (Nvar-0-0) -- (1.6, 0);
\draw[-, dashed] (Nvar-0-1) -- (1.6, -1);
\draw[-, dashed] (Nvar-2-0) -- (2.6, 0);
\draw[-, dashed] (Nvar-2-1) -- (2.6, -1);
\end{tikzpicture}
}
\subfigure[Remove a degree-2 check node]{
\label{fig:fg_reduction_d}
\begin{tikzpicture}[yscale=0.45, xscale=1.2, node distance=0.2cm, auto, semithick]
\node[check] (Ncheck-1-0) at (0.0, 0) {};
\node[check] (Ncheck-1-1) at (0.0, -1) {};
\node[var] (Nvar-0-0) at (0.5, 0) {};
\node[var] (Nvar-0-1) at (0.5, -1) {};
\node[check] (Ncheck-0-0) at (1.0, 0) {};
\node[check] (Ncheck-0-1) at (1.0, -1) {};
\path (Nvar-0-0) edge[-] (Ncheck-1-0);
\path (Nvar-0-1) edge[-] (Ncheck-1-1);
\path (Nvar-0-0) edge[-] (Ncheck-0-0);
\path (Nvar-0-1) edge[-] (Ncheck-0-1);
\path (Nvar-0-1) edge[-] (Ncheck-1-0);
\draw[-, dashed] (Ncheck-0-0) -- (1.3, 0);
\draw[-, dashed] (Ncheck-0-1) -- (1.3, -1);
\draw[-, dashed] (Ncheck-1-1) -- (-0.5, -1);
\draw[->, thick] (1.5, -0.5) -- (1.8, -0.5); 
\node[check] (Ncheck-1-1) at (2.0, -1) {};
\node[var] (Nvar-0-0) at (2.5, 0) {};
\node[check] (Ncheck-0-0) at (3.0, 0) {};
\node[check] (Ncheck-0-1) at (3.0, -1) {};
\path (Nvar-0-0) edge[-] (Ncheck-0-1);
\path (Nvar-0-0) edge[-] (Ncheck-1-1);
\path (Nvar-0-0) edge[-] (Ncheck-0-0);
\draw[-, dashed] (Ncheck-0-0) -- (3.3, 0);
\draw[-, dashed] (Ncheck-0-1) -- (3.3, -1);
\draw[-, dashed] (Ncheck-1-1) -- (1.7, -1);
\end{tikzpicture}
}
\caption{Reducing a polar code sparse factor graph.}
\label{fig:fg_reduction}
\end{figure}

Every pair of degree-2 and degree-3 check nodes in the sparse factor graph is interconnected through a Z-shaped structure shown in Fig.~\ref{fig:fg_reduction_a}, referred to as a \emph{Z-structure}. The possible configurations of frozen variable nodes in a Z-structure are:
\begin{enumerate}
\item[1.] Both the variable nodes on the left are frozen.
\item[2.] Only a single variable node is frozen and due to the channel polarization principle, this must be the degree-1 variable node on the left.
\end{enumerate}
In case (1), we propagate the left frozen variable node values (0's in this case) to the right variable nodes of the Z-structure and eliminate the Z-structure as shown in Fig.~\ref{fig:fg_reduction_b}. 
We note that this step is similar to the one proposed in the simplified successive cancellation decoder~\cite{yazdi_2011} used to reduce the complexity of the SC decoder. We are left with Z-structures in the graph where only a single bit is frozen i.e., the case (2). We reduce such a Z-structure by replacing it with a degree-2 check node as shown in Fig.~\ref{fig:fg_reduction_c}. Now, there are no more frozen variable nodes in the factor graph and hence starting at the code bit variable nodes on the right, we iteratively reduce degree-2 check nodes as shown in Fig.~\ref{fig:fg_reduction_d}.

Next, we show that for LP decoding, the polar code sparse factor graph can be reduced further by eliminating degree-1 auxiliary variable nodes and their check node neighbors from the graph.
\begin{lemma}
The constraints from a parity check node which is connected to a degree-1 auxiliary variable node in the polar code sparse factor graph do not affect the LP decoder solution and hence the degree-1 auxiliary variable node and its check node neighbor can be deleted from the graph. 
\end{lemma} 
\begin{IEEEproof}
From the formulation of LP decoding for polar codes ((\ref{polar_lp}) in Section~\ref{sec:lp_decoding_polar}), we know that the LP decoder objective function is independent of the auxiliary variable nodes (which do not correspond to codeword bits). Hence a degree-1 auxiliary variable node is free to be assigned any feasible value by the LP solver and can be deleted from the graph.    
\end{IEEEproof}   

The steps for reducing a polar code sparse factor graph are described in Algorithm~\ref{algo:FG_reduce}.
Assuming $u_0, u_1, u_2, u_4$ (Fig.~\ref{fig:sparse_fgraph_8}) are the frozen bits, the reduced factor graph (RFG) of a (8, 4) polar code sparse factor graph obtained using Algorithm~\ref{algo:FG_reduce} is illustrated in Fig.~\ref{fig:8_4_reduced_fg}.

\begin{algorithm}
\caption{Reduce Polar Code Sparse Factor Graph}
\label{algo:FG_reduce}
\algsetup{
linenosize=\small
}
\begin{algorithmic}[1]
\REQUIRE Polar code sparse factor graph $\mathbf{H}_\mathcal{P}$, frozen bit indices 
\ENSURE Reduced factor graph 
\STATE \textbf{Step 1:} Propagate frozen variable node pairs as shown in Fig.~\ref{fig:fg_reduction_b} and eliminate the corresponding Z-structures.
\STATE \textbf{Step 2:} Replace Z-structures containing a single frozen variable node with degree-2 check nodes. (Fig.~\ref{fig:fg_reduction_c})
\STATE \textbf{Step 3:} For each degree-2 check node, delete a variable node neighbor connecting all its neighboring check nodes to the other variable node neighbor. (Fig.~\ref{fig:fg_reduction_d})    
\STATE \textbf{Step 4:} Iteratively delete degree-1 auxiliary variable nodes and their check node neighbors until no further degree-1 auxiliary variable nodes exist. 
\end{algorithmic}
\end{algorithm}
\tikzstyle{var} = [circle, fill, minimum size=4pt,inner sep=0pt, outer sep=0pt]
\tikzstyle{check} = [rectangle, fill, minimum size=5pt,inner sep=0pt, outer sep=0pt]
\tikzstyle{selected check} = [check, fill=blue, minimum size=4.5pt]
\tikzstyle{selected var} = [var, fill=blue, minimum size=3.5pt]
\tikzstyle{connect} = [circle, draw, minimum size=2pt, inner sep=-1pt]
\tikzstyle{every node}=[font=\footnotesize]
\begin{figure}
\begin{center}
\begin{tikzpicture}[yscale=0.25, xscale=0.4, node distance=0.2, auto, semithick]
    \node[var] (Nvar-0) [label=below:$x_{0}$] at (-15.0, 0) {};
    \node[var] (Nvar-1) [label=below:$x_{1}$] at (-13.0, 0) {};
    \node[var] (Nvar-2) [label=below:$x_{2}$] at (-11.0, 0) {};
    \node[var] (Nvar-3) [label=below:$x_{3}$] at (-9.0, 0) {};
    \node[var] (Nvar-4) [label=below:$x_{4}$] at (-7.0, 0) {};
    \node[var] (Nvar-5) [label=below:$x_{5}$] at (-5.0, 0) {};
    \node[var] (Nvar-6) [label=below:$x_{6}$] at (-3.0, 0) {};
    \node[var] (Nvar-7) [label=below:$x_{7}$] at (-1.0, 0) {};
    \node[var] (Nvar-8) [label=above:$x_{8}$] at (-8.0, 4.5) {};
    \node[var] (Nvar-9) [label=below:$x_{9}$] at (-7.0, -4.5) {};
    
    \node[check] (Ncheck-0) at (-14.0, 2) {};
	\node[check] (Ncheck-1) at (-10.0, 2) {};
	\node[check] (Ncheck-2) at (-6.0, 2) {};
	\node[check] (Ncheck-3) at (-2.0, 2) {};
	\node[check] (Ncheck-4) at (-11.0, -3) {};
	\node[check] (Ncheck-5) at (-3.0, -3) {};
        
    \path (Nvar-0) edge[-] (Ncheck-0);
    \path (Nvar-1) edge[-] (Ncheck-0);
    \path (Nvar-2) edge[-] (Ncheck-1);
    \path (Nvar-3) edge[-] (Ncheck-1);
    \path (Nvar-4) edge[-] (Ncheck-2);
    \path (Nvar-5) edge[-] (Ncheck-2);
    \path (Nvar-6) edge[-] (Ncheck-3);
    \path (Nvar-7) edge[-] (Ncheck-3);
    \path (Nvar-8) edge[-] (Ncheck-0);
    \path (Nvar-8) edge[-] (Ncheck-1);
    \path (Nvar-8) edge[-] (Ncheck-2);
    \path (Nvar-8) edge[-] (Ncheck-3);
    \path (Nvar-1) edge[-] (Ncheck-4);
    \path (Nvar-3) edge[-] (Ncheck-4);
    \path (Nvar-5) edge[-] (Ncheck-5);
    \path (Nvar-7) edge[-] (Ncheck-5);
    \path (Nvar-9) edge[-] (Ncheck-4);
    \path (Nvar-9) edge[-] (Ncheck-5);
\end{tikzpicture}
\end{center}
\caption{Reduced factor graph (RFG) of a (8, 4) polar code.}
\label{fig:8_4_reduced_fg}
\end{figure}
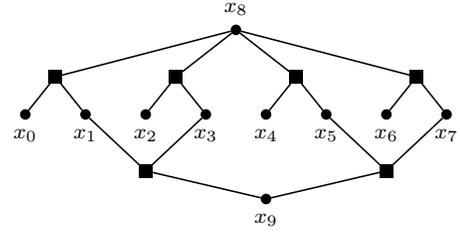
Let $\mathbf{H}_\mathcal{R}$ be a parity-check matrix representation of the polar code reduced factor graph. We show that $\mathbf{H}_\mathcal{R}$ has small degree check nodes necessary to represent the LP polytope efficiently.
\vspace*{-1.2em}
\begin{lemma}
A polar code reduced factor graph $\mathbf{H}_\mathcal{R}$ consists of only degree-3 check nodes.
\end{lemma}
\begin{IEEEproof}
We refer to the steps of Algorithm~\ref{algo:FG_reduce} for this proof. A polar code sparse factor graph has only \mbox{degree-2} and \mbox{degree-3} check nodes. Step~1 deletes check nodes 
in Z-structures with two frozen bits (Fig.~\ref{fig:fg_reduction_b}) and hence does not change the degree of any other check nodes. Step~2 operates on Z-structures with a single frozen bit (Fig.~\ref{fig:fg_reduction_c}) and deletes the \mbox{degree-2} check node while reducing the degree of the \mbox{degree-3} check node by 1. Step~3 iteratively deletes \mbox{degree-2} check nodes and does not affect the \mbox{degree-3} check nodes. Hence, we are left with only \mbox{degree-3} check nodes in the factor graph. Step~4 deletes the \mbox{degree-1} auxiliary variable nodes and their check node neighbors. Therefore, the reduced factor graph $\mathbf{H}_\mathcal{R}$ consists of only \mbox{degree-3} check nodes.   
\end{IEEEproof}

\begin{theorem}
Let $\mathcal{R}$ be the fundamental polytope of the reduced factor graph $\mathbf{H}_\mathcal{R}$ of a polar code $\mathcal{C}$. Then, $\mathcal{R}~\subset~[0, 1]^{d}$,~~$\textrm{where}~d = f(N, r)$ is the dimension of the vectors in $\mathcal{R}$ and is a function of the polar code blocklength $N$ and the rate $r$. Let $\mathcal{P}$ be the polytope obtained from the original sparse factor graph and the frozen bit information of $\mathcal{C}$ and let $\tilde{\mathcal{P}}$ be the projected polytope obtained from the projection of vectors in $\mathcal{P}$ onto the $d$ variables in $\mathbf{H}_\mathcal{R}$. Then,
\begin{IEEEeqnarray}{C}
\mathcal{R} = \tilde{\mathcal{P}}
\end{IEEEeqnarray}
\end{theorem}
\begin{IEEEproof}
First we show that $\tilde{\mathcal{P}} \subseteq \mathcal{R}$ i.e., every vector in polytope $\tilde{\mathcal{P}}$ is also in polytope $\mathcal{R}$. Consider a vector $\mathbf{u} \in \mathcal{P}$; its projection of length $d$, $\tilde{\mathbf{u}} \in \tilde{\mathcal{P}}$, can be constructed by deleting the components of $\mathbf{u}$ corresponding to the variable nodes deleted in Algorithm~\ref{algo:FG_reduce}. It is clear that no step in Algorithm~\ref{algo:FG_reduce} requires a change in the value of a variable node which is not deleted and hence $\tilde{\mathbf{u}} \in \mathcal{R}$.
Next, we show that $\mathcal{R} \subseteq \tilde{\mathcal{P}}$. Let $\mathbf{v} \in \mathcal{R}$; then $\mathbf{v}$ satisfies all the parity checks in $\mathbf{H}_\mathcal{R}$. From the proof of Lemma~2, we know that there are degree-3 parity checks without frozen variable node neighbors in the original sparse factor graph which cannot be reduced. In $\mathbf{H}_\mathcal{R}$, even though the variable node indices participating in these checks may be different from those in $\mathbf{H}_\mathcal{P}$, the parity checks remain unchanged because there is one representative variable node for a group of deleted variable nodes which were constrained to take on the same values. Hence, the set of parity checks in $\mathbf{H}_\mathcal{R}$ is a subset of the parity checks in $\mathbf{H}_\mathcal{P}$ and $\mathbf{v} \in \tilde{\mathcal{P}}$. Therefore, $\mathcal{R} = \tilde{\mathcal{P}}$.
\end{IEEEproof}

LP decoding on the polytope $\mathcal{P}$ has the ML-certificate property~\cite{goela2010} and from Theorem~1 it follows that the polytope $\mathcal{R}$ also has the ML-certificate property. 
We replace the matrix $\mathbf{H}_\mathcal{P}$ with $\mathbf{H}_\mathcal{R}$ in the ACG-ALP-Polar decoder (Algorithm~\ref{algo:ACG_ALP_polar_algo}). The size of the matrix $\mathbf{H}_\mathcal{R}$ is strictly smaller than that of $\mathbf{H}_\mathcal{P}$ for any polar code of rate~$<$~1. Hence the decoding time complexity of the ACG-ALP-Polar decoder can only decrease by using the reduced factor graph.   

\subsection{Simulation Results}
\begin{figure}[!ht]
\includegraphics[width=0.48\textwidth]{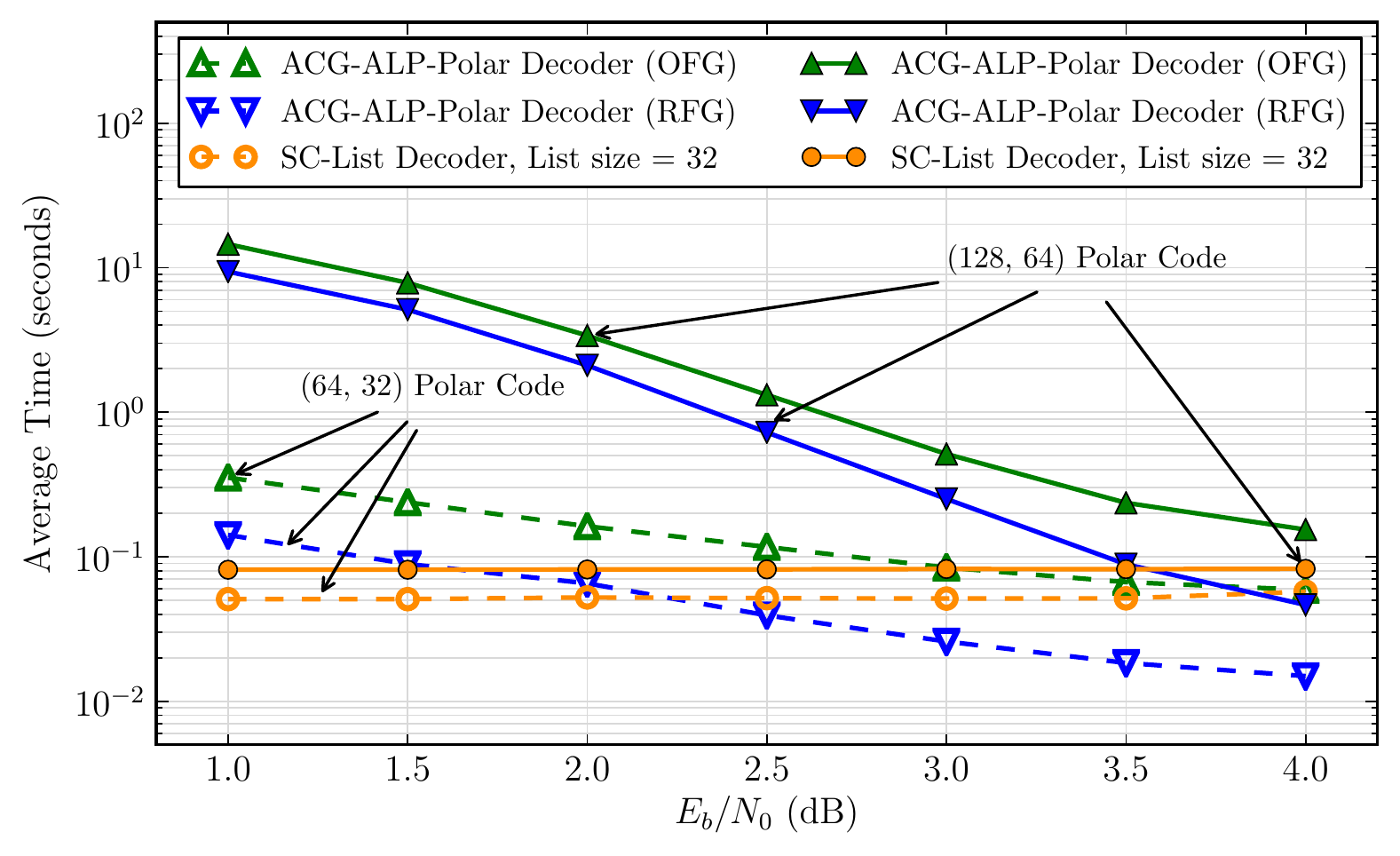}
\setlength{\abovecaptionskip}{-5pt}
\caption{Average time for decoding one codeword of a (64,~32) and (128,~64) polar code over a BAWGN channel. OFG~--~with original factor graph; RFG~--~with reduced factor graph.}
\label{fig:acg_alp_rfg_avgtime}
\end{figure}

\begin{figure}
\includegraphics[width=0.48\textwidth]{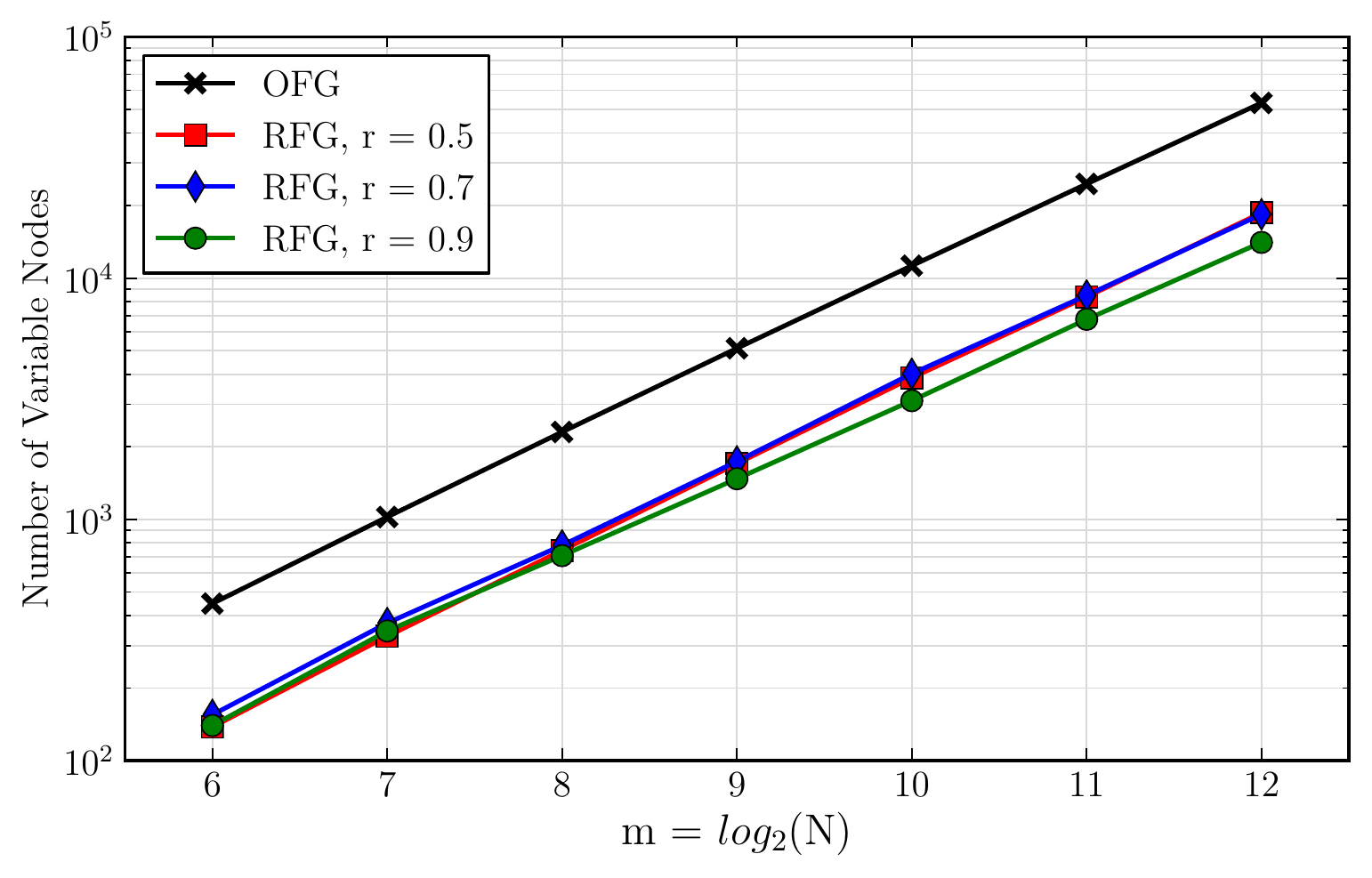}
\setlength{\abovecaptionskip}{-5pt}
\caption{Representation complexity ($d~=~f(N, r)$) for polar codes using the original sparse factor graph (OFG) and the reduced factor graph (RFG).}
\label{fig:ofg_rfg_variable_nodes}
\end{figure}

We present simulation results using the reduced factor graph representation in the ACG-ALP-Polar decoder for the two polar codes discussed in Section~\ref{subsec:acg_alp_polar_sim_results}. 
The FER performance is unchanged (Fig.~\ref{fig:acg_alp_ofg_64} and Fig.~\ref{fig:acg_alp_ofg_128}) and hence is not shown. However, as Fig.~\ref{fig:acg_alp_rfg_avgtime} shows, the decoding time complexity is decreased when using the reduced factor graph representation. 
Compared to the SC-List-32 decoder, the ACG-ALP-Polar decoder with the reduced factor graph has a lower average decoding time complexity at higher SNRs. The reduction in the representation complexity of polar codes using the reduced factor graph is shown in Fig~\ref{fig:ofg_rfg_variable_nodes}.

%% ===========================================
%% Conclusion
%% ===========================================
\section{Conclusion}
\label{sec:conclusion}
We proposed modifications to the ACG-ALP decoder~\cite{zhang2012} which make it suitable for decoding short to moderate length polar codes with FER performance close to the ML performance. 
This indicates that with the proper polytope representation, LP decoding works well for polar codes over a BAWGNC. We also presented an algorithm to generate an efficient reduced factor graph representation of a polar code. This reduced factor graph decreases the decoding time complexity of the ACG-ALP-Polar decoder without degrading its error rate performance. 

\section*{Acknowledgment}
The authors would like to thank Aman Bhatia for helpful discussions. This work was supported in part by the Center for Magnetic Recording Research at the University of California, San Diego and the Western Digital Corporation.

\ifCLASSOPTIONcaptionsoff
  \newpage
\fi

\bibliographystyle{IEEEtran}
\bibliography{references/references}

\end{document}